\documentclass{elsart}
\usepackage{amsfonts,amsmath,array}
\usepackage{amssymb}
\usepackage{graphicx}
\usepackage{bm}

\newcommand {\ee}{\end{equation}}
\newcommand{\be}{\begin{equation}}
\newcommand {\eea}{\end{eqnarray}}
\newcommand{\bea}{\begin{eqnarray}}
\newcommand {\eeas}{\end{eqnarray*}}
\newcommand{\beas}{\begin{eqnarray*}}

\def\fr#1#2{\frac{#1}{#2}}
\def\Fr#1#2{\left(\frac{#1}{#2}\right)}
\def\half{\fr{1}{2}}

\def\Total#1#2{\frac{d#1}{d#2}}
\def\ddt#1{\Total{#1}{t}}
\def\Abs#1{\left|#1\right|}
\def\Re{\mathop{\rm Re}\nolimits}
\def\Im{\mathop{\rm Im}\nolimits}
\def\e{\epsilon}
\def\d{\delta}
\def\l{\lambda}
\def\n{\eta}
\def\a{\alpha}
\def\b{\beta}
\def\g{\gamma}
\def\s{\sigma}
\def\v{\bm}
\def\va{{\v a}}
\def\vu{{\v u}}
\def\vx{{\v x}}
\def\vnu{{\v\nu}}
\def\ph{\varphi}
\def\<{\langle}
\def\>{\rangle}
\def\sgn{\mathop{\rm sgn}\nolimits}
\def\T#1{\widetilde{#1}}
\let\cal\mathcal

\begin{document}

\journal{Physica D}

\bibliographystyle{physicad}

\begin{frontmatter}

\title{Links between dissipation, intermittency, and
helicity in the GOY model revisited}
\author{John C. Bowman}
\address{Department of Mathematical and Statistical Sciences,
University of Alberta, Edmonton, Alberta T6G 2G1, Canada}
\author{Charles R. Doering} 
\address{Department of Mathematics, University of Michigan,
Ann Arbor, MI 48109-1043, USA}
\address{Michigan Center for Theoretical Physics, Ann Arbor, 
MI 48109-1120, USA}
\author{Bruno Eckhardt}
\address{Fachbereich Physik, Philipps-Universit\"at Marburg, 
D-35032 Marburg, Germany}
\address{IREAP, IPST and Burgers Program, University of Maryland, 
College Park, MD 20754, USA}
\author{Jahanshah Davoudi} 
\address{Fachbereich Physik, Philipps-Universit\"at Marburg, 
D-35032 Marburg, Germany}
\author{Malcolm Roberts}
\address{Department of Mathematical and Statistical Sciences,
University of Alberta, Edmonton, Alberta T6G 2G1, Canada}
\author{J\"org Schumacher}\footnote{Present address: Department of
Mechanical Engineering, Technische Universit\"at Ilmenau, P.O. Box 100565
D-98684 Ilmenau, Germany.}
\address{Fachbereich Physik, Philipps-Universit\"at Marburg, 
D-35032 Marburg, Germany}

\bigskip
\flushleft{Submitted to Physica D: 18 October 2005\\
Revised: 16 March 2006}
\pagebreak[4]

\begin{abstract}
High-resolution simulations within the GOY shell model are used to study
various scaling relations for turbulence. A power-law relation between the
second-order intermittency correction and the crossover from the inertial
to the dissipation range is confirmed. Evidence is found for the
intermediate viscous dissipation range proposed by Frisch and Vergassola. It is
emphasized that insufficient dissipation-range resolution
systematically drives the energy spectrum towards
statistical-mechanical equipartition. In fully resolved simulations the
inertial-range scaling exponents depend on both model parameters; in
particular, there is no evidence that the conservation of a helicity-like
quantity leads to universal exponents.
\end{abstract}

\begin{keyword}

GOY shell model \sep dissipation scale \sep intermittency \sep helicity 
\sep intermediate dissipation range \sep turbulence
\PACS{47.27.Eq; 47.52.+j}

\end{keyword}

\end{frontmatter}

\section{Introduction}
One of the most challenging aspects of turbulent fluid flow is the rapid 
rise in the number of active degrees of freedom as the Reynolds number 
increases.  Given present day computational resources,
this renders direct numerical simulation of the asymptotic 
high-Reynolds-number regime of fully developed turbulence intractable.
Various reduced models of the dynamics have therefore been developed to
capture the essential features of the turbulent cascade of energy from
large to small scales
\cite{Gled,YO_a,YO_b,YO_c,OY,siggia,eggers_1991,Eggers,DynSyst,Luca}. 
These models reflect
many of the statistical features expected to influence the energy
transfer in realistic flows~\cite{siggia,gl94a,gl94b,glr96}. 
They have been used to study intermittency, multifractal
behavior, and the scaling of energy and dissipation 
(see also
\cite{bof2000a,bof2000b}). They have also been extended
to include passive scalars~\cite{Jensen_92}, magnetic fields
\cite{Biskamp_94}, and even polymers~\cite{Benzi_04}. 

In this note we focus on the GOY shell model, so named after
its initial proponents Gledzer~\cite{Gled} and 
Ohkitani and Yamada~\cite{YO_a,YO_b,YO_c} (see also \cite{DynSyst,Luca} for
reviews). We consider 
within high-resolution, fully resolved, and converged numerical simulations 
four related issues: the Reynolds number dependence of the
crossover to the dissipation scale; the existence of an intermediate
dissipation range; the effect of modal truncation
in the viscous range on inertial-range scaling; and the
relation between inertial-range scaling and the presence
of a second conserved helicity-like quantity.
Some of these aspects have been discussed for shell models in isolation
before (e.g.~\cite{SL_95,bottle_falk,bottle_lmg,bottle_lmg_long}).
Key observations of the present paper are: the crossover from
inertial to dissipative scales takes place at a
length much smaller than the Kolmogorov
estimate; these scales must be properly resolved to
avoid contaminating the inertial range; and, finally, the high-resolution,
fully converged simulations presented here do
not support the conjecture~\cite{kadanoff1} that inertial-range exponents
in the helicity-preserving GOY model are universal.

The usual estimate for the scale on which dissipation begins
to dominate, based on the energy dissipation
density $\epsilon$ and the kinematic viscosity~$\nu$, is
the Kolmogorov scale
\be
\n=\left(\frac{\nu^3}{\epsilon}\right)^{1/4}.\label{K41}
\ee
However, numerical simulations of passive scalars show that
in order to recover small-scale fluctuations, 
gradients, and dissipation accurately,
one has to go well below this scale~\cite{joerg}.
A straightforward line of reasoning expanded on in 
Sect.~\ref{kd} shows that in the presence of intermittency 
corrections, as reflected in the energy
density scaling like $k^{-5/3-\delta}$ with $0<\delta < 4/3$,
the dissipative range sets in at a scale $\n_d$ that is a 
fraction of $\n$ that
decreases as the Reynolds number increases (i.e.\ as the viscosity decreases):
\be
\n_d=\n \left(\fr{\n}{L}\right)^{3\delta/(4-3\delta)}
\sim \nu^{1/(\fr{4}{3}-\d)}.\label{etad0}
\ee
The integral or outer scale $L$ appearing here is the characteristic scale
of the external stirring force.
A more detailed argument for the variation
of the energy density in the transition between the inertial and viscous
ranges, based on the multifractal model of intermittency, 
shows that an {\it intermediate dissipation range\/} develops that widens with
increasing Reynolds number~\cite{FV}. In Sect.~\ref{idr},
we present high-resolution numerical evidence for the existence of an
intermediate dissipation range in the GOY model. However, this part of the
spectrum is so steep that it contributes negligibly to the energy balances
used in Sect.~\ref{kd} to determine the dissipation wavenumber. 

As part of these studies we also considered the influence of 
modal truncations on spectra. If the modes are truncated
very far in the viscous range, it is reasonable to expect that the
behavior will not change. However, as the cutoff moves closer and closer
to the dissipation scale, it begins to inhibit sufficient dissipation. As a
consequence, the pileup of energy close to the transition between
the inertial- and viscous-dominated regimes 
(the {\it bottleneck effect\/}, which also occurs in
simulations of the Navier--Stokes equation
\cite{bottle_falk,bottle_lmg,bottle_lmg_long}) is enhanced to the point where
the system tends towards a weakly damped equilibrium:
the distribution of energy approaches a $1/k$ scaling, corresponding to an
equipartition of the energy contents of the geometrically spaced shells.
A similar kind of effect has been noted in models with
hyperviscosity, where the rapid quenching of high-wave number
modes also effects the inertial-range scaling~\cite{SL_95}.

In Sect.~\ref{helicity}, using fully resolved simulations, we
revisit the studies of Kadanoff {\it et al.\/}~\cite{kadanoff1,kadanoff2}. 
These authors considered
different parameters for the GOY model and reported that 
the intermittency corrections fall into two categories, 
depending on whether or not the model preserves a helicity-like
quadratic quantity. On revisiting these studies with higher-resolution
simulations, we do not find such a correspondence,
but rather a persistent dependence of the intermittency corrections
on all parameters of the model.

Before addressing these issues, it is helpful to describe in the next
section the underlying model and some related technical details.

\section{Preliminaries}
The GOY model has, as primary
dynamical variables, complex velocities~$u_n$ representing the
velocity field in shell number $n$, where $n=0,1,\ldots,N-1$. They evolve
according to
\be
\left(\ddt{} + \nu k_n^2\right) u_n = {\cal S}_n+ F_n,\label{GOY}
\ee
where
\be
{\cal S}_n=i k_n \left(\alpha u_{n+1}^*u_{n+2}^*
+\frac{\beta}{\lambda} u_{n-1}^*u_{n+1}^* + 
\frac{\gamma}{\lambda^2} u_{n-1}^*u_{n-2}^*\right),
\ee
given the boundary conditions $u_{-2}=u_{-1}=u_{N}=u_{N+1}=0$.
The wavenumbers $k_n=k_0\lambda^n$ scale geometrically.
It is customary to rescale time so that $\a=1$ and require that
$\a+\b+\g=0$, so that the nonlinear terms in 
$S_n$ conserve the energy $\half\sum_n |u_n|^2$.
A second invariant $\half\sum_n k_n^p|u_n|^2$ is also conserved,
where $p=-\log_\l(-\b-1)$. Of particular interest is the case where
$\lambda=1/(1+\beta)$, when this invariant takes the form 
of a quantity $H=\half\sum_n (-1)^n k_n |u_n|^2$ with the same dimensions
and sign indefiniteness as the helicity invariant of three-dimensional
Navier Stokes turbulence.

For the standard case considered by Kadanoff {\it et al.}
\cite{kadanoff1,kadanoff2},
$k_0=2^{-3}$, $\lambda=2$, $\beta=\gamma=-1/2$, $\nu=10^{-7}$, $N=22$ and
the forcing is confined to shell $3$:
\be
F_n=f\delta_{n,3}.
\ee
The amplitude of the external forcing is constant: $f=5(1+i)\times 10^{-3}$.
In these models, one observes an inertial-range power-law scaling
of the mean shell energy $\half \langle |u_n|^2\rangle 
\sim k_n^{-2/3}$, corresponding to the
energy spectrum
$E(k_n)=\half \<|u_n|^2\>/(k_{n+1}-k_n)\sim k_n^{-5/3}$
and reminiscent of the ``K41'' scaling predicted by the Kolmogorov
theory~\cite{k41}. Here $\<\>$ denotes an ensemble average.
At higher shell indices one observes an exponential decrease of the shell
amplitudes, corresponding to a viscous range~\cite{nelkin}. 

On multiplying~(\ref{GOY}) by $u_n^*$ and summing over shells $m$ to~$N-1$,
one arrives at the energy balance
\be
\ddt{}\sum_{n=m}^{N-1} \Abs{\<u_n\>}^2=\Pi_m-\epsilon_m,\label{Pibalance}
\ee
where
$\Pi_m=2\Re \sum_{n=m}^{N-1} \<{\cal S}_n u_n^*\>$
represents the transfer of energy by the nonlinear terms into shells
$m$ to~$N-1$ and
\be
\epsilon_m=2\nu \sum_{n=m}^{N-1} k_n^2 \<\Abs{u_n}^2\>-
2\Re\sum_{n=m}^{N-1} \<F_n u_n^*\>,
\ee
is the total transfer, {\it via\/} dissipation and forcing, {\it out\/} of
this range of shells.
A positive value for $\Pi_m$ represents a flow of
energy to shells $m$ and higher. When $\nu=0$ and $F=0$, energy conservation
implies that 
\be
\Pi_m=-2\sum_{n=0}^{m-1} \<S_n u_n^*\>,
\ee
so that $\Pi_0=\Pi_N=0$.
In a statistical steady state, the left-hand side of~(\ref{Pibalance})
vanishes and $\Pi_m=\epsilon_m$. This condition serves as an excellent
numerical diagnostic for discerning when a steady state has been reached.
In practice, appealing to the ergodic theorem, running time averages (say
from $T/2$ to the final time~$T$) are used to approximate the required ensemble
averages.

As pointed out in~\cite{Pisarenko93}, it is highly advantageous to use an
exponential integrator to solve~(\ref{GOY}) exactly on the
linear time scale. Exponential
integrators~\cite{Certaine60,Hochbruck98,Beylkin98,Cox02,Hochbruck05} are
well suited to linearly stiff systems that exhibit dynamics on a
wide range of time scales. They allow for much larger time steps than
either classical nonstiff methods or simple integrator factor (Rosenbrock)
methods~\cite{Hairer91}. However, instead of using the second-order
exponential Adams--Bashforth scheme used by 
Pisarenko {\it et al.\/}~\cite{Pisarenko93} or the fixed time step schemes of
Cox and Matthews~\cite{Cox02}, we used the exponential
version~\cite{Bowman05} of the adaptive third-order
Bogacki--Shampine~\cite{Bogacki89} Runge--Kutta integrator described in
Appendix~\ref{E_RK3}. With a nonstochastic forcing, this scheme can reuse
the final source evaluation from the previous time step.

Instead of a constant forcing, it is also possible to apply a central
white-noise stochastic forcing, where $f$ is delta-correlated in time.
The advantage of this forcing is that it yields a
prescribed mean energy injection $\epsilon=\half\langle \Abs{f}^2 \rangle$,
independent of the lattice size and forcing~\cite{Novikov64}.

\begin{figure}[ht]
\begin{center}
\includegraphics{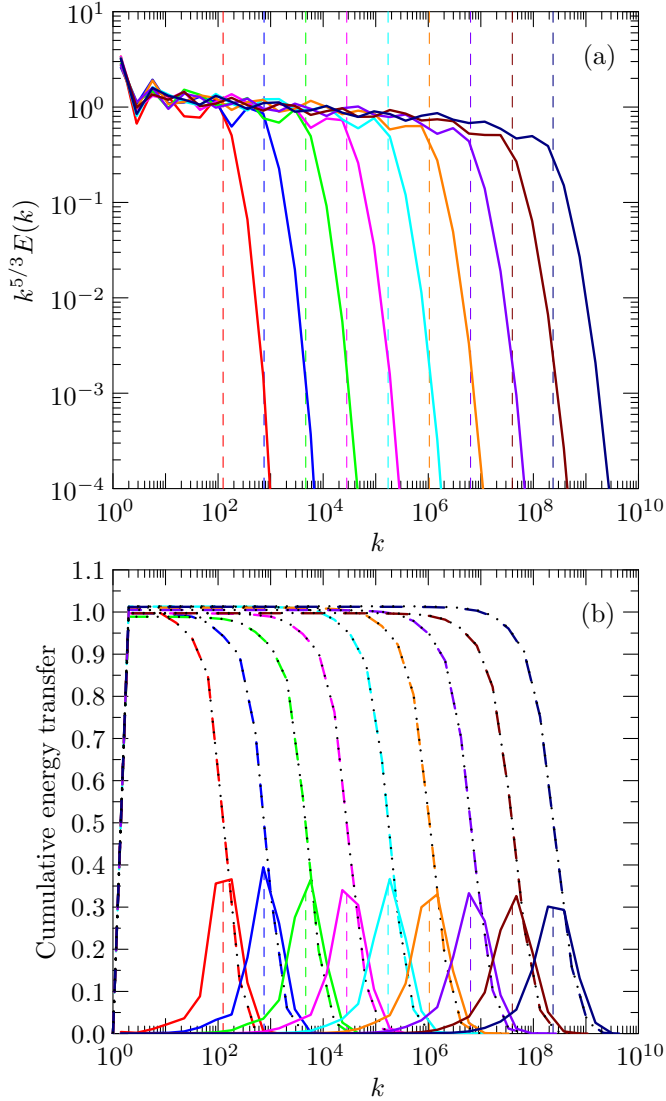}
\caption{(a) Compensated energy densities for the
GOY model driven by a white-noise random forcing, normalized by the $K41$
law, for  $\nu=10^{-3}$, $10^{-4}$, $10^{-5}$, $10^{-6}$, $10^{-7}$,
$10^{-8}$, $10^{-9}$, $10^{-10}$, and $10^{-11}$.
(b) Cumulative energy transfer functions: the dashed (dotted) curves
indicate the value of $\Pi_n$ ($\e_n$) at the wavenumbers $k_n$.
The energy dissipation $\nu k^2 \langle\left|u_n\right|^2\rangle$ (solid
curves) is maximal very near $k_d$ (vertical dashed lines).
}\label{z0kd}
\end{center}
\end{figure}

The fully resolved spectra and cumulative energy transfers
$\Pi_n$ and $\epsilon_n$ corresponding to a series of decreasing values
of $\nu$ are shown in Fig.~\ref{z0kd}. 
Here we set $k_0=1$ and used the standard values $(\beta,\lambda)=(-1/2,2)$
with a complex-valued white-noise random forcing 
restricted to the first shell,
such that the total energy injection $\epsilon$ is~$1$.
Initially all shells were assigned the same nonzero real amplitude
(a statistical mechanical equipartition of energy);
the complex forcing and nonlinearity then cause them to develop imaginary
parts. The dashed vertical lines in Fig.~\ref{z0kd} separate regions of equal
energy dissipation, providing a convenient definition for the
dissipation wavenumber~$k_d$; namely, $2\nu\int_0^{k_d}k^2E(k)\,dk=\e/2$.
As seen by the near coincidence of $k_d$ with the maxima of the
energy dissipation, this definition of $k_d$ closely approximates a point
of inflection, with respect to shell number, of the cumulative energy
transfer \cite{Leveque01}.

\begin{figure}[ht]
\begin{center}
\includegraphics{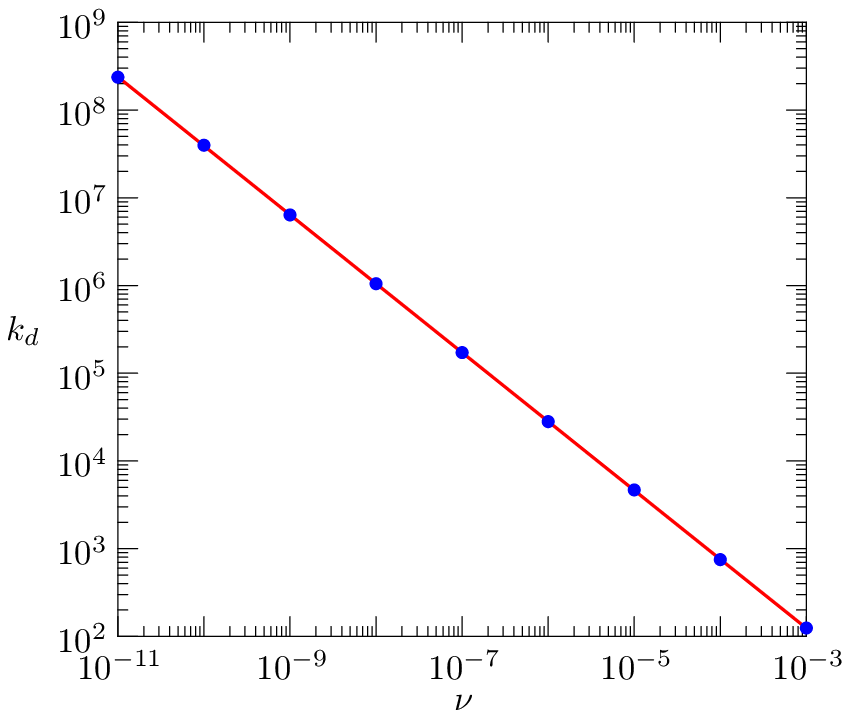}
\caption{Scaling of $k_d$ vs.\ $\nu$ for the simulations in Fig.~\ref{z0kd}.
The line joining the data points is a least-squares fit.}\label{z0kdvnu}
\end{center}
\end{figure}

The highest resolution case was run $1.2\times 10^9$ adaptive time steps
for a total of $9756$ time units, or roughly $4000$ large-eddy turnover
times.  In Fig.~\ref{z0kdvnu}, we illustrate the power-law behavior of
$k_d$ vs.\ $\nu$ evident in Fig.~\ref{z0kd}; using a least-squares fit, we
find that \be k_d\sim \nu^{-0.7855},\label{kdvnufit} \ee with the exponent
determined to within a statistical error of $0.0005$.  Thus, instead of the
K41 value $-0.75$ arising from~(\ref{K41}), we obtain an exponent much
closer to the anomalous value $-0.775$ deduced from~(\ref{etad0}), using
the second-order intermittency correction $\delta=0.0438$ measured in
section~\ref{kd}.

The slow decrease with wavenumber of the energy spectra compensated by the
simple algebraic scaling indicates an intermittency correction to the
exponent. The relationship between this intermittency
correction and the dissipation wavenumber $k_d$ will be considered
in the next section.

\section{Intermittency-length scale paradox}
\label{kd}

Deviations of the $p$-th order structure function scaling exponents
from the Kolmogorov value $p/3$ require the presence of a
length scale $\ell$ for dimensional correctness:
\be
S_p(k_n)=\left<\Abs{u_n}^p\right>\sim 
\left(\fr{\e}{k_n}\right)^{p/3} (k_n\ell)^{-\delta_p}
\sim\e^{p/3} k_n^{-\zeta_p} \ell^{-\delta_p},
\ee
where $\zeta_p=p/3+\d_p$. Two natural choices for this length scale
are the external, stirring scale $L$ and the small scale $\n_d$
at which dissipation terminates the inertial range.
Concerning this choice, we now give two elementary arguments that at first
glance lead to seemingly conflicting results.

The energy spectrum may be expressed in terms of the intermittency exponent
$\d=\d_2$:
\be
E(k)=C \e^{2/3} k^{-5/3} (k\ell)^{-\d}.\label{spectrum}
\ee
On balancing the energy injection $\e$ and energy dissipation between
the largest scale $L$ and some dissipation scale $\n_d$, one finds
\bea
\e&=&2\nu\int_{2\pi/L}^{2\pi/\n_d} k^2 E(k)\,dk
=2C\nu\e^{2/3} \ell^{-\d} \int_{2\pi/L}^{2\pi/\n_d} k^{1/3-\d}\,dk\nonumber\\
&\sim& \nu\e^{2/3} \ell^{-\d} \Fr{1}{\n_d}^{4/3-\d}
\qquad(\n_d \ll L),\label{Edissbalance}
\eea
where we have tacitly assumed that $\d < 4/3$, so that the upper
integration limit dominates. 
If one takes $\n_d$ to be the Kolmogorov scale $\n=(\nu^3/\e)^{1/4}$,
then on comparing the left- and right-hand sides of~(\ref{Edissbalance}), one
obtains
\be
\e\sim \nu\e^{2/3} \Fr{\n}{\ell}^\d \fr{\e^{1/3}}{\nu}.
\label{DoeringEckhardt}
\ee
For $\d \neq 0$, such a balance is possible only if $\ell\sim \n$; 
that is, the relevant choice for~$\ell$ is the dissipation scale.

On the other hand, the Cauchy--Schwarz inequality implies that
structure functions satisfy~\cite{Frisch95}
\be
S_{\fr{p}{2}+\fr{q}{2}}\le S_{p}^{1/2} S_{q}^{1/2}.
\ee
On expressing $S_p\sim (\e r)^{p/3}(r/\ell)^{\d_p}$, where $r=2\pi/k$,
one obtains a convexity condition on $\d_p$:
\be
\Fr{r}{\ell}^{\d_\fr{p+q}{2}} \le \Fr{r}{\ell}^{\fr{\d_p}{2}+\fr{\d_q}{2}},
\ee
or
\be
1 \le \Fr{r}{\ell}^{\fr{\d_p+\d_q}{2}-\d_{\fr{p+q}{2}}}.
\ee
So if $\ell$ is the smallest excited length scale ($\ell\sim\n$), then
$r/\ell > 1$ for all~$r$, which would imply  that $\d_p$ (and hence
$\zeta_p$) is convex. Alternatively, if $\ell$ is the largest excited
length scale ($\ell\sim L$), then $r/\ell < 1$ for all $r$, and
$\d_p$ (and hence $\zeta_p$) is concave.

\begin{figure}[ht]
\begin{center}
\includegraphics{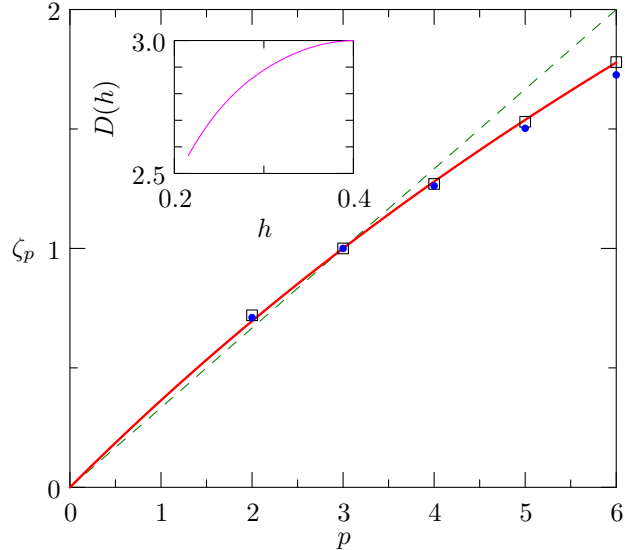}
\caption{Concavity of structure functions exponents $\zeta_p$ vs.\ $p$
for $\beta=-0.5$ and $\lambda=2$. The circles represent numerical values
obtained by a least-squares fit of the flux~(\ref{flux}) between $k=10^2$ and
$10^6$ for the case $\nu=10^{-11}$. The dashed line is the Kolmogorov
prediction $\zeta_p=p/3$; the solid line indicates the She--L\'ev\^eque scaling
$\zeta_p=p/9+2-2(2/3)^{p/3}$, and the squares represent the experimental values
measured for 3D turbulence by van de Water \& Herweijer~\cite{HvdW_95}.
In the inset we graph the dimension $D(h)$ determined from our numerical
$\zeta_p$ values by inversion of the multifractal Legendre transformation.}
\label{z0zetavp}
\end{center}
\end{figure}

\begin{table}
\begin{center}
\begin{tabular}{|>{$}c<{$} || >{$}c<{$} | >{$}c<{$} | >{$}c<{$} | >{$}c<{$} | >{$}c<{$} | >{$}c<{$} | }
\hline
p&2&3&4&5&6\\
\hline
\zeta_p&\phantom{\pm}0.7105&\phantom{\pm}0.9999&\phantom{\pm}1.262
&\phantom{\pm}1.503&\phantom{\pm}1.726\\
\hline
\Delta_{\zeta_p}&\pm0.0005&\pm0.0005&\pm0.001&\pm0.001&\pm0.002\\
\hline
\end{tabular}
\caption{Numerically computed structure function exponents~$\zeta_p$ and
the statistical errors~$\Delta_{\zeta_p}$ (from a least-squares fit
between $k=10^2$ and $10^6$) for $\beta=-0.5$ and $\lambda=2$.}
\label{z0zetavalues}
\end{center}
\end{table}

Both Navier--Stokes turbulence and the GOY model have concave exponents,
as illustrated in Fig.~\ref{z0zetavp} and tabulated in
Table~\ref{z0zetavalues} for $(\beta=-1/2,\lambda=2)$. In fact, 
the curvature of $\zeta_p$ is very close to that observed
experimentally for three-dimensional Navier--Stokes
turbulence~\cite{HvdW_95}. This supports $\ell=L$ (see also
Ref.~\cite{Esser99}), even though~(\ref{DoeringEckhardt}) appears to
require that $\ell=\n$. The above calculations assume spectra with a sharp
cut-off at the dissipation scale, but the same analysis with the
same conclusion can be performed for spectra with an 
exponential cut-off. 

This apparent contradiction is easily resolved: a steeper-than-Kolmogorov
spectrum necessitates integrating to scales smaller than $\n$ to obtain
sufficient dissipation~\cite{FV}. If we set $\n_d\sim \n \Fr{\n}{L}^{\sigma}$,
we obtain the energy balance
\be
1\sim \Fr{\n}{\ell}^\d  \Fr{\n}{L}^{\s(\d-4/3)}.\label{energybalance}
\ee
This is consistent with $\ell=L$ if $\d+\s(\d-4/3)=0$, that is, if
\be
\s^{-1}=\fr{4}{3}\d^{-1}-1.
\ee
For $0 < \d < 4/3$, we see that $\s > 0$. The conclusion is that
the inertial range extends to a scale $\n_d$ that is smaller than the
Kolmogorov scale $\n$:
\be
\n_d\sim \n \Fr{\n}{L}^{\sigma}
\sim \n^{1+\s}=\Fr{\nu^3}{\e}^{\fr{(1+\s)}{4}},
\ee
for a fixed outer scale $L$.
When the energy injection $\e$ is also fixed, we thus have
\be
\n_d\sim\nu^{1/(\fr{4}{3}-\d)}.\label{etad}
\ee
Equivalently, on setting $\ell=L$, one can obtain that this scaling
directly from~(\ref{Edissbalance}).

To verify this relation numerically, we followed Kadanoff {\it et al.}
\cite{kadanoff2} and
computed $\zeta_p$ {\it via\/} a least-squares fit of the time-averaged
$p$th-order flux
\be
\Sigma_{n,p}=
\left<
\Abs{\Im\left(u_n u_{n+1} u_{n+2}+
\fr{1+\beta}{\lambda}u_{n-1}u_nu_{n+1}\right)}^{p/3}
\right>,\label{flux}
\ee
in order to filter out the well-known period-three oscillation in solutions of
the GOY model~\cite{Pisarenko93}.
We chose the interval $[10^2,10^6]$ for doing the least-squares fit, as 
the logarithmic slopes of $\Sigma_{n,p}$ were nearly flat in this region
and, in the case $n=3$, very close to the exact value of $-1$.
From this fit, we obtained the value $\zeta_2=0.7105\pm 0.0005$
tabulated in Table~\ref{z0zetavalues}. The error quoted here and elsewhere
is the statistical error from the least-squares fit to the data. 
Other significant sources of error are the choice of wavenumber interval
for the fit and the time-averaging window. By varying the fitting interval
and time-averaging window, we confirmed that the error from these sources
has roughly the same magnitude as the statistical fitting errors
(cf. Fig.~\ref{zeta6}). On substituting the measured second-order
intermittency correction $\delta_2=\zeta_2-2/3=0.0438$ into~(\ref{etad}),
we obtain
\be
k_d\sim \nu^{-0.775},\label{kddelta}
\ee
in remarkable agreement with the observed 
dissipation wavenumber scaling~(\ref{kdvnufit}). This result has
implications for theoretical predictions of how the dissipation
wavenumber should depend on Reynolds number. Practically, it suggests
for numerical simulations how the maximum truncation wavenumber should
scale as the viscosity is decreased. Our results in section~\ref{helicity}
indicate, for the standard case of Ref.~\cite{kadanoff1}, that the upper
truncation wavenumber must be chosen to be at least three times larger than the
dissipation wavenumber $k_d$ in order to obtain reliable inertial-range
dynamics.

\section{Intermediate dissipation range}
\label{idr}

Fluctuations in the energy transport and 
dissipation were previously considered by Frisch and Vergassola
within a multifractal model \cite{FV}
and studied numerically by Nakayama \cite{Nakayama01}. Frisch and Vergassola
noted that the inertial-range scaling exponent $h$ for velocity
differences has an influence on the spectrum at the dissipative scale
as well. Equating the viscous time and
the turnover time, they found an $h$-dependent viscous cut-off,
\be
\n(h)\sim\nu^{1/(1+h)}\label{FVetad}.
\ee
This agrees with our equation (\ref{etad}) after the
identification $h=1/3-\delta$.
Frisch and Vergassola appealed to a geometric characterization of
the sets of points that contribute to a certain velocity scaling exponent $h$
for a given separation scale and introduced the Hausdorff dimension $D(h)$
of these sets. In the present shell model, such an interpretation is not
available; nevertheless, the spectrum can still be expressed in terms of
an effective dimension
\begin{equation}
D(h)=\inf_p (ph+3-\zeta_p)
\end{equation}
calculated directly from the scaling exponents $\zeta_p$, 
as displayed in the inset of Fig.~\ref{z0zetavp}.
The dominating scaling exponent $h$ is given by the slope of the
graph of $\zeta_p$ at $p=2$ in Fig.~\ref{z0zetavp}. On fitting a Bezier
cubic spline through the numerical data points, we find that $h=0.309$,
which yields 
\be
k_d\sim \nu^{-0.764}.
\ee
In comparison with the measured scaling~(\ref{kdvnufit}), this result is
slightly less accurate than~(\ref{kddelta}).
If we use the She--Leveque expression \cite{SL_94} for $\zeta_p$ instead
of our measured values, we find $h=0.317$, which yields the dissipation
wavenumber scaling exponent $-0.759$.

\begin{figure}[ht]
\begin{center}
\includegraphics{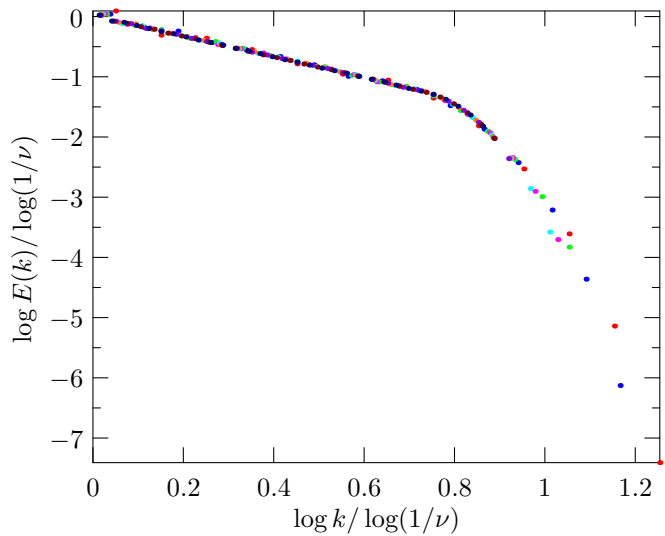}
\caption{Collapse of the spectra in Fig.~\ref{z0kd} with the rescaling
proposed by Frisch and Vergassola.}
\label{z0multifractal}
\end{center}
\end{figure}

\begin{figure}[ht]
\begin{center}
\includegraphics{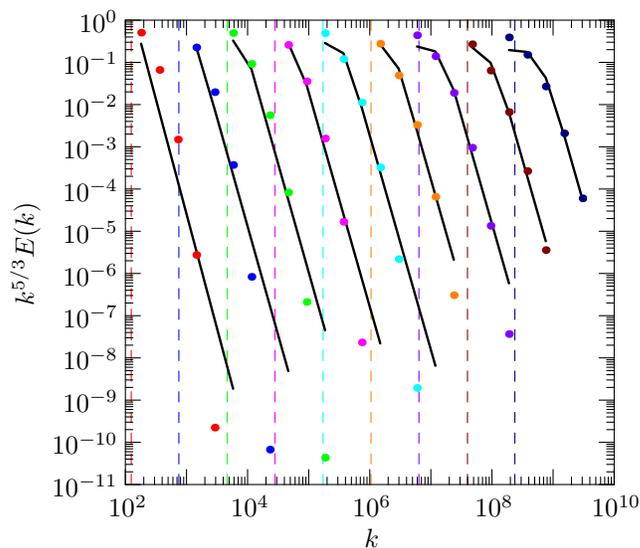}
\caption{Intermediate dissipation ranges in the simulations
of Fig.~\ref{z0kd}. For each case, the dots indicate the portion of the
numerical spectrum for $k > 1/\n$, the solid line depicts the spectrum
given by~(\ref{FVintermed}), and the vertical dashed line separates
regions of equal energy dissipation.
}\label{z0ekvk53intermed}
\end{center}
\end{figure}

\begin{figure}[ht]
\begin{center}
\includegraphics{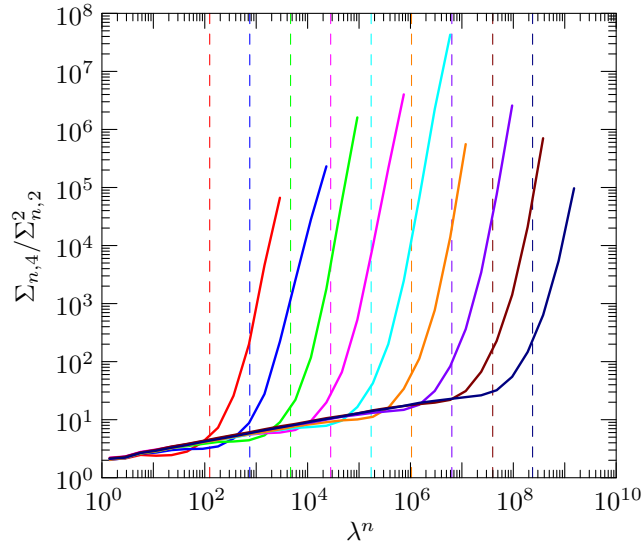}
\caption{The rapid growth of the flux kurtosis for the simulations of
 Fig.~\ref{z0kd} in each intermediate dissipation range, beginning at
$k_d$ (denoted by a vertical dashed line).
}\label{kurtosis}
\end{center}
\end{figure}

We show in Fig.~\ref{z0multifractal} that the rescaling function proposed
by Frisch and Vergassola collapses all of the spectra in Fig.~\ref{z0kd} to
a single curve.
In Fig.~\ref{z0ekvk53intermed}, we use our computed function $D(h)$ to
provide explicit numerical evidence for the existence of intermediate
dissipation ranges in these simulations. We compare the
portion of each spectrum in Fig.~\ref{z0kd} for $k \ge 1/\n$ with the
intermediate dissipation range spectrum proposed by Frisch and
Vergassola~\cite{FV}:
\be
E(k)=A k^{-4-2h(k)+D(h(k))} \qquad (k \ge 1/\n)\label{FVintermed},
\ee
where $h(k)=-1-\log\nu/\log k$. We set the constant of proportionality
$A$ to $0.5$. For the lower resolution runs, one notices only a narrow
intermediate range, with the numerical spectra quickly entering the full
dissipation range. At the highest resolutions, the intermediate
dissipation range appears to consist of all resolved wavenumbers above
$1/\n$. Since the energy in these intermediate dissipation ranges decays
very rapidly beyond $k_d$, it contributes negligibly to the energy
balance~(\ref{energybalance}). In Fig.~\ref{kurtosis}, we illustrate the
rapid growth of intermittency in the intermediate dissipation range
by plotting the flux kurtosis $\Sigma_{n,4}/\Sigma_{n,2}^2$ vs. $\lambda^n$,
consistent with the findings of Ref.~\cite{Chevillard05},
although a direct comparison of our deterministic model with the random
cascade model studied there is not available.
Similar (but less-smooth, due to the period-three oscillation) results were
obtained for the kurtosis of the shell velocities $u_n$. 

In summary, we were able to determine the effective dimension $D(h)$ and
study its effects on the dissipation-range energy spectrum. The
steepening of the energy spectrum in the inertial range implies
that the dissipation range sets in well after the Kolmogorov scale is
reached, namely at the scale determined by (\ref{etad}) or
(\ref{FVetad}). Frisch and Vergassola \cite{FV} note that this reduction in
scale compared to the Kolmogorov scale is relatively small, but our numerical
simulations are precise enough to resolve it.

\section{Dissipation-scale effects on inertial-range dynamics}
\label{helicity}

\begin{figure}[ht]
\begin{center}
\includegraphics{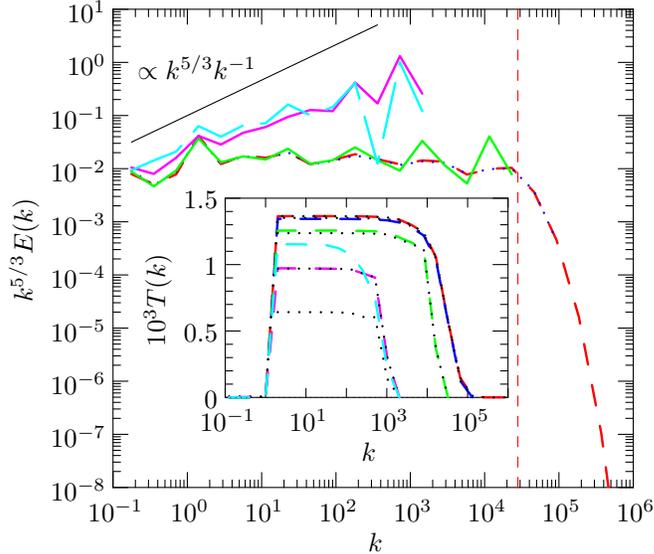}
\caption{The effect of a finite number of shells on the spectrum for the
standard case $(\beta=-1/2,\lambda=2)$ of Ref.~\cite{kadanoff1}.
For $N=26$ (dashed) and $N=20$ (dotted),
there is no discernible difference between the energy spectra, showing that a
sufficient number of shells in the viscous subrange is included. 
The vertical dashed line denotes the dissipation wavenumber $k_d$.
For $N=18$ (solid), the dissipation range is barely resolved and the
inertial range is contaminated. For $N=14$, the inertial range tends toward the
statistical equipartition spectrum $k^{-1}$: the short solid curve indicates
the spectrum obtained with a conservative integrator; the long-dashed curve
indicates the spectrum obtained with a nonconservative exponential
integrator that violates global energy balance.}
\label{Kadanoff53inset}
\end{center}
\end{figure}

\begin{figure}[ht]
\begin{center}
\includegraphics{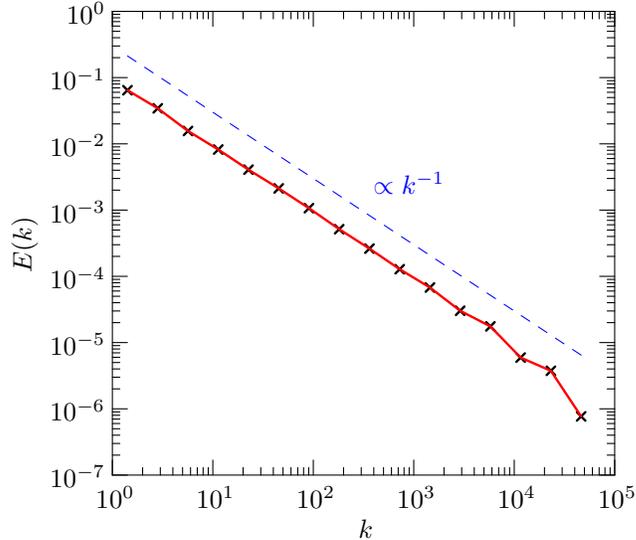}
\caption{Equipartition spectrum for 16 shells obtained with the
conservative C--RK5 integrator with $\epsilon=\nu=0$.}
\label{equipartition}
\end{center}
\end{figure}

When the number of shells is kept fixed, and the viscosity is reduced,
the viscous range will become narrower and narrower, until it finally
disappears. Sch\"orghofer {\it et al.}~\cite{kadanoff2} 
call this the regime of {\it strong interaction}. However, as this happens, 
the effect of the lack of resolution no longer remains confined to 
the small scales: it begins to affect the inertial-range intermittency
corrections as well. This effect is illustrated in Fig.~\ref{Kadanoff53inset}
for simulations with fixed viscosity and varying $N$. 
In particular, we see on comparing the $N=18$ and $N=20$ spectra
that it is necessary to resolve the small-scale dynamics well beyond the
dissipation wavenumber $k_d$ (denoted by the vertical dashed line).
The smallest number of shells for which the large-scale
dynamics was adequately resolved was $N=20$, which requires a maximum
wavenumber more than three times higher than $k_d$.
For the case $N=14$, where the viscous range was completely resolved, it
was necessary to use the conservative C--RK5 integrator 
described in Appendix~\ref{C--RK5} to discretize the nonlinear terms in a
manner that respects exact energy conservation. Otherwise, as 
illustrated for the nonconservative solution in the transfer function
inset, viscous dissipation will be insufficient to absorb the temporal
energy discretization error, so that a global energy balance can never be
reached. In Fig.~\ref{equipartition}, we use this conservative integrator
to illustrate the eventual equipartition of shells energies and resulting
$k^{-1}$ energy spectrum, in the extreme limit of zero dissipation,
given an initial $k^{-2}$ spectrum.

\begin{figure}[ht]
\begin{center}
\includegraphics{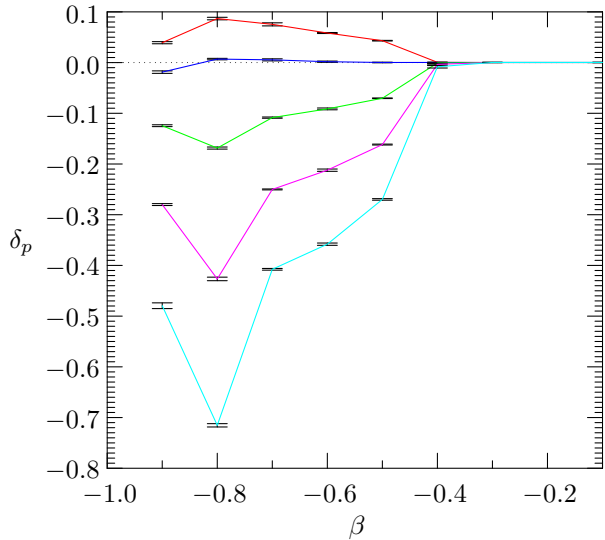}
\caption{Scaling anomalies~$\d_p$, with $\lambda=2$, for
$p=2$, $3$, $4$, $5$, and $6$ (from top to bottom) vs.\ $\beta$, as
obtained by a least-squares fit between $k=500$ and $2\times 10^6$. The
number of shells is $32$ and the viscosity $\nu$ is $10^{-11}$. The system is
driven with a stochastic white-noise force on the first shell such that
the energy injection $\epsilon=1$.}
\label{dzetavbetal2}
\end{center}
\end{figure}

In view of these simulations, the conjecture in~\cite{kadanoff2}
on the role of preservation of the helicity $H$
has to be reconsidered. Kadanoff {\it et al.}\ noted that their simulations
for the standard case $(\beta=-1/2,\lambda=2)$ closely reproduced 
experimentally measured intermittency exponents of 
three-dimensional turbulence (cf. Fig.~\ref{z0zetavp}). At the time, it had
already been observed~\cite{BLLP_95} that for fixed $\lambda$, the
intermittency corrections strongly depend on the parameter $\beta$, as
illustrated in Fig.~\ref{dzetavbetal2}. 
Kadanoff {\it et al.\/} then hypothesized that experimentally
observed exponents should be obtained on the curve
$\lambda=1/(1+\beta)$ along which the second conserved invariant of the GOY
model is analogous to the helicity invariant of three-dimensional
Navier--Stokes turbulence.

\begin{figure}[ht]
\begin{center}
\includegraphics{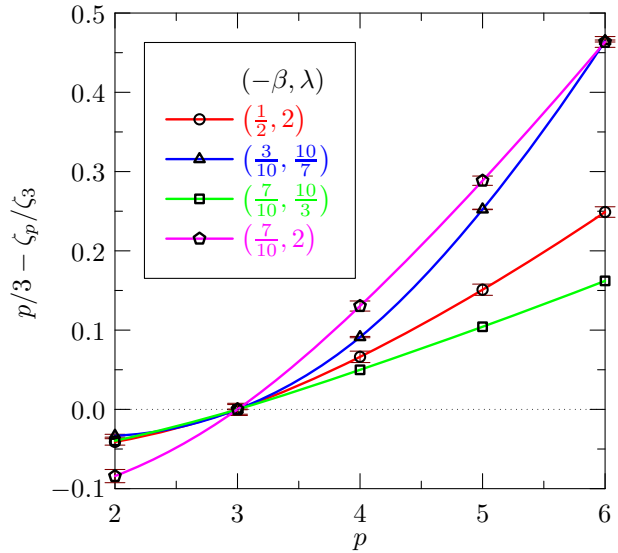}
\caption{Recalculation of Fig.~4 from Kadanoff {\it et al.},
fitting between $k=20$ and $200$.
The first three $(-\beta,\lambda)$ pairs respect helicity conservation; the
fourth does not. All four cases exhibit large anomalies, independent of
whether helicity is preserved.}
\label{Kadanoff4}
\end{center}
\end{figure}

\begin{figure}[ht]
\begin{center}
\includegraphics{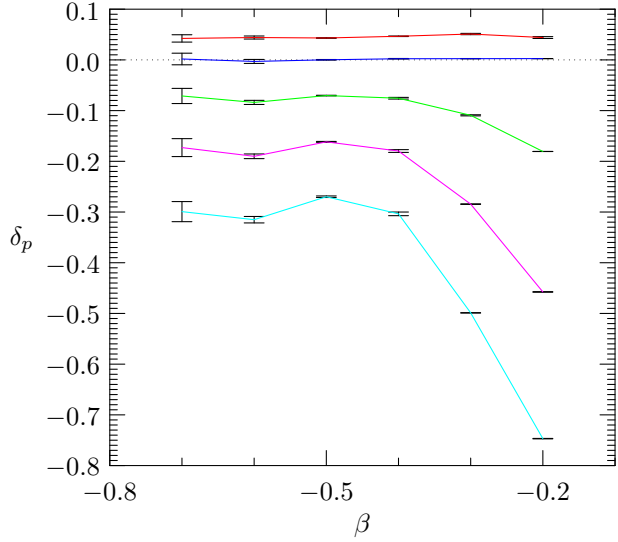}
\caption{
Scaling anomalies~$\d_p$ from top to bottom for $p=2$, $3$, $4$, $5$,
and $6$ in the helicity-preserving case where $\lambda=1/(1+\beta)$ for
the values of $\beta$ and number of shells $N$ listed in
Table~\ref{helicityparam}, as obtained by a least-squares fit between
$k=500$ and $2\times 10^6$. The viscosity $\nu$ is $10^{-11}$. The system is
driven with a stochastic white-noise force on the first shell such that
the energy injection $\epsilon=1$.}\label{dzetavbetaH}
\end{center}
\end{figure}

\begin{figure}[ht]
\begin{center}
\includegraphics{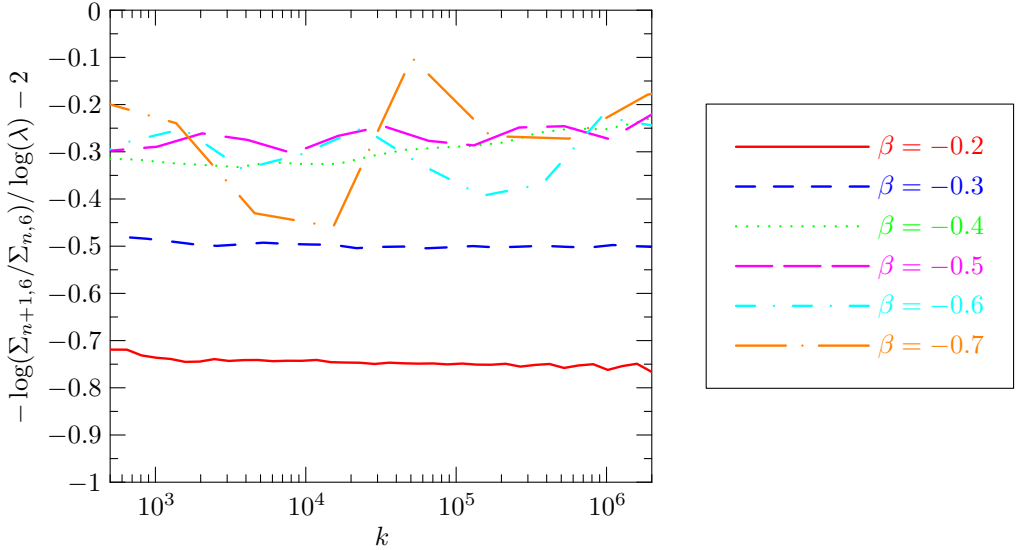}
\caption{Determination of $\d_6$ from the logarithmic slope of
  $\Sigma_{n,6}$ for different values of $\beta$, with $\lambda=1/(1+\beta)$.}
\label{zeta6}
\end{center}
\end{figure}

However, on repeating their simulation with a sufficient number of shells,
(particularly the case with $\beta=-3/10$ and $\lambda=10/7$, which requires
$44$ rather than $22$ shells), we find that the structure function exponents
change dramatically. In particular, in view of Fig.~\ref{Kadanoff4}, it
appears that the distinction made between the cases with and without
helicity conservation in Fig.~4 of~Ref.~\cite{kadanoff1} cannot be upheld:
helicity preservation does not lead to unique intermittency corrections.
This is further emphasized in Fig.~\ref{dzetavbetaH}, where the variation
of the anomalies $\d_p$ with $\beta$ along the helicity preserving curve is
shown. Fig.~\ref{zeta6} emphasizes that the computed slopes are well
resolved, and independent of the fitting interval, particularly for small
values of $\beta$ (corresponding to small $\lambda$ and large $N$),
where the departure from the conjecture is seen to be most significant.

\begin{figure}[ht]
\begin{center}
\includegraphics{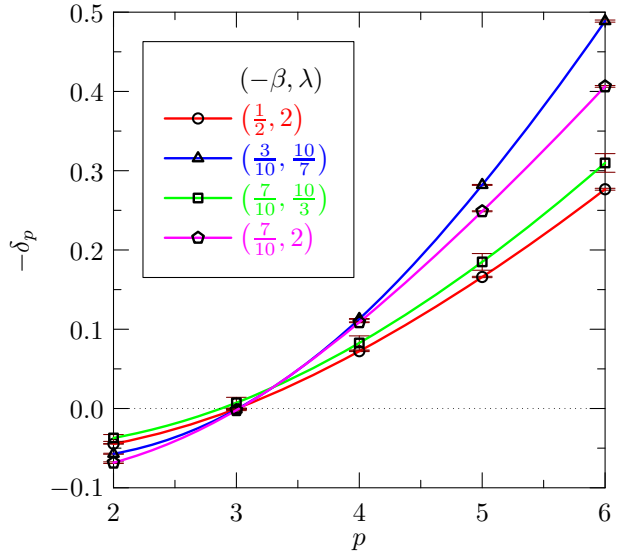}
\caption{Analog of Fig.~\ref{Kadanoff4} generated from the high-resolution
data in Figs.~\ref{dzetavbetal2} and~\ref{dzetavbetaH}.}
\label{z0dzetvp}
\end{center}
\end{figure}

While Fig.~\ref{Kadanoff4} might appear to support the conjecture
at least for $p=2$, we see that neither
Fig.~\ref{dzetavbetaH} nor the higher-resolution analog Fig.~\ref{z0dzetvp}
of Fig.~\ref{Kadanoff4} generated using data from Figs.~\ref{dzetavbetal2}
and~\ref{dzetavbetaH}, agree with it. Our
high-resolution numerical data set is sufficiently well resolved that we
did not need to invoke extended self-similarity to obtain the results in
Fig.~\ref{z0dzetvp}.

\begin{table}
\begin{center}
\begin{tabular}{|>{$}c<{$} || >{$}c<{$} | >{$}c<{$} | >{$}c<{$} | >{$}c<{$} | >{$}c<{$} | >{$}c<{$} | }
\hline
\beta&	-\fr{2}{10} & -\fr{3}{10} & -\fr{4}{10} & -\fr{1}{2} & -\fr{6}{10}
& -\fr{7}{10}\\
\hline
\lambda&\fr{10}{8}&\fr{10}{7}&\fr{10}{6}&2&\fr{10}{4}&\fr{10}{3}\\
\hline
N&99&62&43&32&24&19\\
\hline
\delta_2&0.04&0.05&0.047&0.043&0.044&0.042\\
\hline
\delta_4&-0.18&-0.11&-0.08&-0.07&-0.08&-0.07\\
\hline
\delta_6&-0.75&-0.50&-0.30&-0.27&-0.32&-0.30\\
\hline
\end{tabular}
\caption{Values of $\beta$, $\lambda$, $N$, $\delta_2$, $\delta_4$, and
$\delta_6$ for the helicity-preserving cases.}
\label{helicityparam}
\end{center}
\end{table}

\section{Conclusions}

In this work, we exploited the numerical advantages of the GOY model,
combined with efficient integration algorithms, to obtain fully
converged spectra that allowed us to address links between dissipation,
scaling, and conserved quantities.

We documented that the typical dissipation scale is much
smaller than the Kolmogorov length, due to the inherent
steepening of the spectrum by intermittency corrections. While the
detailed form of the spectrum beyond this point is influenced
by the probability distribution of the velocity scaling exponent, as discussed
by Frisch and Vergassola \cite{FV}, the delayed transition is not.
Overall, the effect is relatively small, but easily identified in 
simulations of the GOY model.

The shell model simulations presented here also underscore the need
to fully resolve the small scales. They demonstrate how insufficient
resolution of the viscous subrange can affect inertial-range scaling
exponents. In particular, when the viscous range is fully
resolved, one observes significant variations in the spectral exponents even
along the helicity preserving curve,
thus disproving a previous conjecture \cite{kadanoff1}. 
For the standard case of Ref.~\cite{kadanoff1}, we found that the upper
truncation wavenumber must be at least three times higher than the
dissipation wavenumber $k_d$ in order to capture the proper inertial-range
dynamics.

While current direct numerical simulations of three-dimensional
Navier--Stokes turbulence barely reach into the inertial range, the
implication of this relation between the second-order intermittency
correction and the dissipation wavenumber is that, for future
numerical simulations, it will become increasingly important
to account properly for unresolved viscous scales.

{\bf Acknowledgements}

We thank the Alexander von Humboldt Foundation, the Deutsche
Forschungsgemeinschaft, the National Science and Engineering Research
Council of Canada, the Pacific Institute for Mathematical Sciences,
and the U.S. National Science Foundation for their generous support. This
work was completed while BE was visiting the University of Maryland as
Burgers Professor. It is a pleasure to thank the Burgers board and in
particular Dan Lathrop for their hospitality at the University of
Maryland. We also thank Sigfried Grossmann for discussions on the 
contents of section III and Andrew Hammerlindl and Tom Prince for
their contributions to {\tt Asymptote} ({\tt asymptote.sf.net}), the
descriptive vector graphics language used to draw the figures in this work.

\appendix
\section{Exponential adaptive (3,2) Bogacki--Shampine Runge--Kutta integrator}
\label{E_RK3}
The GOY model can be written as a set of ordinary differential equations
for the components of the vector $\vu=(u_0,u_1,\ldots,u_{N-1})$:
\be
\ddt\vu+\vnu\vu=\v S(\vu),\label{ODE}
\ee
where $\vnu$ is a diagonal matrix with entries
$(\nu k_0^2,\nu k_1^2,\ldots,\nu k_{N-1}^2)$.

A general $s$-stage autonomous Runge--Kutta scheme that evolves an initial
value $\vu_0$ by a time step $\tau$ to a final value $\vu_s$ may be written in
the form
\be
\vu_i=\vu_0+\tau\sum_{j=0}^{i-1}\va_{ij}\v S(\vu_j)\qquad (i=1,\ldots,s).
\label{s-stage}
\ee

Letting $\vx=-\tau\vnu$, the $4$-stage exponential Bogacki--Shampine
Runge--Kutta order (3,2) embedded pair~\cite{Bowman05} uses the coefficients
\beas
\va_{10}&=&\half\ph_1\left(\fr{1}{2}\vx\right),\nonumber\\
\va_{20}&=&\fr{3}{4}\ph_1\left(\fr{3}{4}\vx\right)-a_{21},\ \ \ 
\va_{21}=\fr{9}{8}\ph_2\left(\fr{3}{4}\vx\right)+
\fr{3}{8}\ph_2\left(\fr{1}{2}\vx\right),\nonumber\\
\va_{30}&=&\ph_1(\vx)-\va_{31}-\va_{32},\ 
\va_{31}=\fr{1}{3}\ph_1(\vx),\ \ 
\va_{32}=\fr{4}{3}\ph_2(\vx)-\fr{2}{9}\ph_1(\vx),\nonumber\\
\va_{40}&=&\ph_1(\vx)-\fr{17}{12}\ph_2(\vx),\ \ 
\va_{41}=\fr{1}{2}\ph_2(\vx),\ \ 
\va_{42}=\fr{2}{3}\ph_2(\vx),\ \ 
\va_{43}=\fr{1}{4}\ph_2(\vx),
\eeas
where the (here diagonal) matrix functions $\ph_1$ and $\ph_2$
are given by $\ph_1(\vx)=\vx^{-1}(e^\vx-\v 1)$ and
$\ph_2(\vx)=\vx^{-2}(e^\vx-\v 1-\vx)$.
The value $\vu_3$ is the third-order solution, while the value $\vu_4$
provides a second-order estimate that can be used to optimally adjust the
time step. 
Since $\v S(\vu_3)$ is just the value of $\v S$ at the initial $\vu_0$ for
the next time step, no additional source evaluation is actually required to
compute $\vu_4$ (unless an extra stochastic forcing term is added to
$\vu_3$ before proceeding to the next time step).

Care must be exercised when evaluating the diagonal components $\ph_1(x)$
and $\ph_2(x)$ near $0$; optimized double precision routines for evaluating
these functions are available at 
{\tt www.math.ualberta.ca/$\sim$bowman/phi.h}.

\section{Conservative adaptive (5,4) Runge--Kutta integrator}
\label{C--RK5}
Here we describe a fifth-order conservative integrator (C--RK5) similar to the
second-order conservative predictor--corrector algorithm introduced
in~\cite{Bowman97b,Shadwick99,Kotovych02}.
We first consider the case where each component of $\vu$
in~(\ref{ODE}) is real. The scheme shares the same first five
stages as the classical $6$-stage Cash--Karp Fehlberg Runge--Kutta
(4,5) embedded (RK5) pair. However, the final conservative stage is
derived by applying 
the final conventional stage to the evolution equation for each
(say,~$r^{\rm th}$) component $u$ of $\vu$:
\be
\ddt{u^2}=2\T S(\vu)u,
\ee
where $\T S(\vu)=\v S_r(\vu)-\vnu_r\vu_r$.
One then transforms the resulting estimate for the new value of $u^2$
back into an estimate for $u$. The conventional RK5 estimate can be used to
resolve the appropriate branch of the square root:
\be
u_6=\sgn\Big(u_0+\tau\sum_{j=0}^5\va_{5j} \T S(\vu_j)\Big)
\sqrt{u_0^2+2\tau\sum_{j=0}^5\va_{5j} \T S(\vu_j)u_j}.\label{C--RK5C}
\ee
In regions where the argument of the square root becomes negative,
the time step needs to be temporarily reduced.
The resulting solution is then compared with the conventional fourth-order
estimate $u_0+\tau\sum_{j=0}^5\va'_{5j} \T S(\vu_j)$ (for a second set of
coefficients $\va'_{5j}$) to adjust the time step appropriately. 
When $u$ is complex, one applies~(\ref{C--RK5C})
separately to the real and imaginary parts of $u$.

The transfer function $\Pi_M$ is based on partial sums of the time averages of 
${\cal S}_n u_n^*$ from time $t_1$ to $t_2$. We compute these time averages as
$[T_n(t_2)-T_n(t_1)]/(t_2-t_1)$, where
\be
\ddt{T_n}={\cal S}_nu_n^*.\label{Tn}
\ee
Energy conservation relies on the fact that $\sum_{n=0}^{N-1}T_n$ is invariant
(since $\sum_{n=0}^{N-1}{\cal S}_nu_n^*=0$). 
Since integrators of the form~(\ref{s-stage}) conserve such invariants that are
linear in the evolved variable~\cite{Kotovych02}, one can use a
conventional RK5 integrator to conservatively evolve~(\ref{Tn}).

We compute time averages of $u_n^2$ as $[M_n(t_2)-M_n(t_1)]/(t_2-t_1)$, where
\be
\ddt{M_n}=\Abs{u_n}^2.\label{Mn}
\ee
Energy conservation implies that $\sum_{n=0}^{N-1}M_n$ should grow linearly
with time; hence we must use conservative values of $u_n$ on the
right-hand side of~\ref{Mn}. Since we only compute conservative solutions
$u_n$ at the beginning and the end of each time step, we are restricted to
using a (second-order) trapezoidal rule to perform the integration
of~(\ref{Mn}).

\end{document}